\documentclass[aps,pre,groupedaddress,longbibliography]{revtex4-1}

\usepackage{graphicx}
\usepackage{float}
\usepackage{color}
\usepackage{amsmath}
\usepackage[linktocpage,colorlinks=true,linkcolor=blue,citecolor=blue,breaklinks=true]{hyperref}
\usepackage{verbatim}
\usepackage{bm}
\renewcommand{\vec}[1]{\bm{\mathrm{#1}}}
\newcommand{\dd}{\,\mathrm{d}}
\newcommand{\micron}[0]{\,$\mu$m}
\newcommand{\supS}[0]{^\mathrm{S}} 
\newcommand{\supP}[0]{^\mathrm{P}} 

\begin{document}

\preprint{}

\title{Entrainment and scattering in microswimmer--colloid interactions}

\author{Henry Shum}
\affiliation{Department of Applied Mathematics, University of Waterloo, Waterloo, Ontario N2L 3G1, Canada}
\email[]{henry.shum@uwaterloo.ca}

\author{Julia M. Yeomans}
\affiliation{The Rudolf Peierls Centre for Theoretical Physics, University of Oxford, Oxford, OX1 3NP, UK}
\email[]{julia.yeomans@physics.ox.ac.uk}
\homepage[]{http://www-thphys.physics.ox.ac.uk/people/JuliaYeomans/}


\date{\today}

\begin{abstract}
{
We use boundary element simulations to study the interaction of model microswimmers with a neutrally buoyant spherical particle. The ratio of the size of the particle to that of the swimmer is varied from $R\supP / R\supS \ll 1$, corresponding to swimmer--tracer scattering, to $R\supP / R\supS \gg 1$, approximately equivalent to the swimmer interacting with a fixed, flat surface. We find that details of the swimmer and particle trajectories vary for different swimmers. However, the overall characteristics of the scattering event fall into two regimes, depending on the relative magnitudes of the impact parameter, $\rho$, and the collision radius, $R^\mathrm{coll}=R\supP + R\supS$. The range of particle motion, defined as the maximum distance between two points on the trajectory, has only a weak dependence on the impact parameter when $\rho < R^\mathrm{coll}$ and decreases with the radius of the particle. In contrast, when $\rho>R^\mathrm{coll}$ the range decreases as a power law in $\rho$ and is insensitive to the size of the particle. We also demonstrate that large particles can cause swimmers to be deflected through large angles. In some instances, this swimmer deflection can lead to larger net displacements of the particle. Based on these results, we estimate the effective diffusivity of a particle in a dilute bath of swimmers and show that there is a non-monotonic dependence on particle radius. Similarly, we show that the effective diffusivity of a swimmer scattering in a suspension of particles varies non-monotonically with particle radius.
}
\end{abstract}



\maketitle

\section{Introduction}
There is considerable current interest in the physics of microswimming. This is due to the relevance of bacterial swimming as an example of self-propulsion at low Reynolds number~\cite{purcell_life_1976}, and also because of the technological and medical applications that are promised by the possibility of designing robust, steerable swimming micro-robots~\cite{zhang_artificial_2009, sitti_miniature_2009, kim_microbiorobotics:_2012, wang_nano/microscale_2012, peyer_bio-inspired_2013, gao_artificial_2015, katuri_designing_2017}. Recent advances in imaging and nanotechnology are allowing experiments investigating the swimming and stirring properties of bacteria~\cite{wu_particle_2000, sokolov_enhanced_2009, drescher_fluid_2011, lauga_dance_2012} and demonstrating the potential for harnessing bacterial motion to perform work, such as driving microscopic gears~\cite{sokolov_swimming_2010, di_leonardo_bacterial_2010}.

Many bacteria and artificial microswimmers move in complex environments where surfaces, colloids and biopolymers influence their motion~\cite{ottemann_roles_1997, azam_sea_2001, wilking_biofilms_2011, gao_artificial_2015}. Experiments have shown that some bacteria (e.g., \textit{Escherichia coli}) can swim near solid surfaces for over a minute before eventually escaping~\cite{drescher_fluid_2011} while other species (e.g., \textit{Vibrio alginolyticus} and \textit{Caulobacter crescentus}) swim away within a second of encountering a surface~\cite{kudo_asymmetric_2005, li_accumulation_2011}. Simple theoretical models of bacterial swimming suggest that hydrodynamic interactions lead to attraction between a swimmer and a nearby planar surface~\cite{berke_hydrodynamic_2008, drescher_fluid_2011}. More detailed hydrodynamic simulations confirm this hydrodynamic trapping effect~\cite{giacche_hydrodynamic_2010, shum_modelling_2010} and also show that it is strongly dependent on details of the cell and flagellum shape and elastic effects~\cite{shum_effects_2012}. For example, it was found that elongated (high aspect ratio) cell bodies and short flagella encourage escape from surfaces whereas low aspect ratio cell bodies and long flagella lead to attraction to surfaces~\cite{shum_modelling_2010}.  

Interestingly, in both \textit{E.\ coli}~\cite{berke_hydrodynamic_2008} and \textit{C.\ crescentus}~\cite{li_accumulation_2011}, the density of bacteria close to the surface is observed to be higher than in the bulk fluid. This accumulation of bacteria at a substrate can lead to attachment and colonization of surfaces. Indeed, bacterial motility has been found to be an important factor in biofilm initiation~\cite{conrad_physics_2012, tuson_bacteriasurface_2013}. 

Other recent experimental and theoretical work has considered the behavior of passive particles in suspensions of swimming bacteria~\cite{wu_particle_2000, jepson_enhanced_2013, kasyap_hydrodynamic_2014} and algae~\cite{leptos_dynamics_2009, kurtuldu_enhancement_2011}. These studies demonstrate that transport of tracers can be significantly enhanced by the activity of swimmers. For example, Wu and Libchaber~\cite{wu_particle_2000} found that passive, spherical particles as large as 10\micron{} in diameter in a bacterial bath exhibited superdiffusive motion on timescales under 1\,s. On long timescales, motion was diffusive but the effective diffusion coefficient was around 100 times larger than that expected from Brownian motion in thermal equilibrium. 

As a step toward understanding the transport of passive particles in a bath of active swimmers, theoretical models have been analyzed for the advection of point-like tracers in the flow field produced by single swimmers~\cite{dunkel_swimmer-tracer_2010, thiffeault_stirring_2010, pushkin_fluid_2013-1, pushkin_fluid_2013, mino_induced_2013, morozov_enhanced_2014, mueller_fluid_2017}. Since neutrally buoyant, self-propelled microswimmers are free of net force and torque, the far field generated by an individual swimmer has the form of a force dipole (or higher order multipole). This flow field causes tracers far from the swimmer to move in (almost closed) loops. A tracer can become entrained when a swimmer passes close to it, leading to large displacements before the tracer escapes from the swimmer's influence. 

Such long distance entrainment, though infrequent, was found to contribute significantly to tracer motion in the presence of the alga, \textit{Chlamydomonas reinhardtii}~\cite{jeanneret_entrainment_2016}. Similar experimental observations have not been reported for bacteria, though theoretical results with model bacteria indicate that long distance entrainment of tracers is possible~\cite{pushkin_fluid_2013-1}. One potential explanation for the difference in observations between bacteria and algae is the different mechanism of propulsion. \textit{C.\ reinhardtii} is a puller, using a pair of flagella at the front of the organism to pull itself forward. In contrast, most bacteria swim with the flagella pushing the cell body from behind. Another important distinction, however, is the size of the organism. The cell body of \textit{C.\ reinhardtii} is approximately spherical with a diameter of 10\micron{}~\cite{leptos_dynamics_2009} whereas an \textit{E.\ coli} cell is typically 2--3\micron{} long and 1\micron{} wide~\cite{wu_particle_2000}. Since tracer particles used in experiments are often at least 1\micron{} in diameter~\cite{mino_enhanced_2011, valeriani_colloids_2011, kasyap_hydrodynamic_2014}, it is not clear that such large particles behave purely as ``tracers'' in the presence of bacteria. For instance, it has been shown that larger particles have shorter entrainment distances when a swimmer collides head on because steric interactions keep the particles further away from the no-slip surface of the swimmer, allowing them to slip around the swimmer and be left behind more quickly~\cite{mathijssen_contact_nodate}.

Nonetheless, several experiments have reported that swimmers cause considerable motion of passive particles even much larger than the swimmers~\cite{wu_particle_2000, sokolov_swimming_2010, di_leonardo_bacterial_2010}. This scenario, in which a swimmer interacts with colloid or particle of finite size, has received little theoretical attention and is the subject of the current numerical study. Our aim is to provide results for model systems which are realistic enough to mirror the behavior of real swimmers, thus providing a benchmark for future experiments. We consider varying colloid sizes, hence showing how to link existing results for a tracer particle (colloid much smaller than the swimmer size) and a swimmer moving near a flat surface (colloid much larger than the swimmer size). We consider three different model swimmers and calculate how the particle and swimmer are scattered in unbounded, three-dimensional space as a function of the impact parameter. This will allow us to discuss the effects of finite particle size on two phenomena: the stochastic motion of a particle in a dilute suspension of swimmers and the dispersion of a swimmer in a colloidal suspension.

In the next sections we describe the model and the scattering geometry. We then present our results, showing that the scattering is highly dependent on the details of the model swimmer. We find that the range through which the colloid moves has only a weak dependence on impact parameter, $\rho$, if this is less than the collision radius. For $\rho$ greater than the collision radius, the range decreases as a power law in $\rho$. We also show that swimmers can be deflected through large angles by large colloids and that some swimmers can become hydrodynamically bound to colloids of large enough radius. The trajectories from simulated scattering events are used to calculate the effective diffusivities of a swimmer in a suspension of particles and of a colloid in a suspension of swimmers. We show that both diffusivities (swimmer and particle) vary non-monotonically with the size of the particle.

\section{Numerical method and model swimmers}

Many of the features of swimmer--particle scattering are dependent on the details of the swimmer considered. Therefore, we choose to compare three different swimmer models. The first is the spherical squirmer, a model for organisms propelled by small body shape deformations proposed by Lighthill~\cite{lighthill_squirming_1952} and adopted in many theoretical studies of microswimmers~\cite{blake_spherical_1971, ishikawa_hydrodynamic_2006, ishikawa_coherent_2008, Pak2014}. Spherical swimmer models are generally more tractable as analytical solutions of Stokes flows can be obtained in simple cases, for example, using bispherical coordinates~\cite{lee_motion_1980}. Relatively complex problems, such as finding the instantaneous velocities of two interacting swimmers, can then be solved analytically or semi-analytically~\cite{sharifi-mood_pair_2016, papavassiliou_exact_2017}. Obtaining trajectories, however, requires numerical integration methods. The other two swimmers we consider resemble bacteria with a stiff, helical, rotating tail such as \textit{Rhodobacter sphaeroides} or \textit{V.\ alginolyticus}. These model bacteria are distinguished by choosing parameters so that they tend to either swim away from, or parallel to, flat surfaces. 

\begin{figure}
 \centering
 \includegraphics{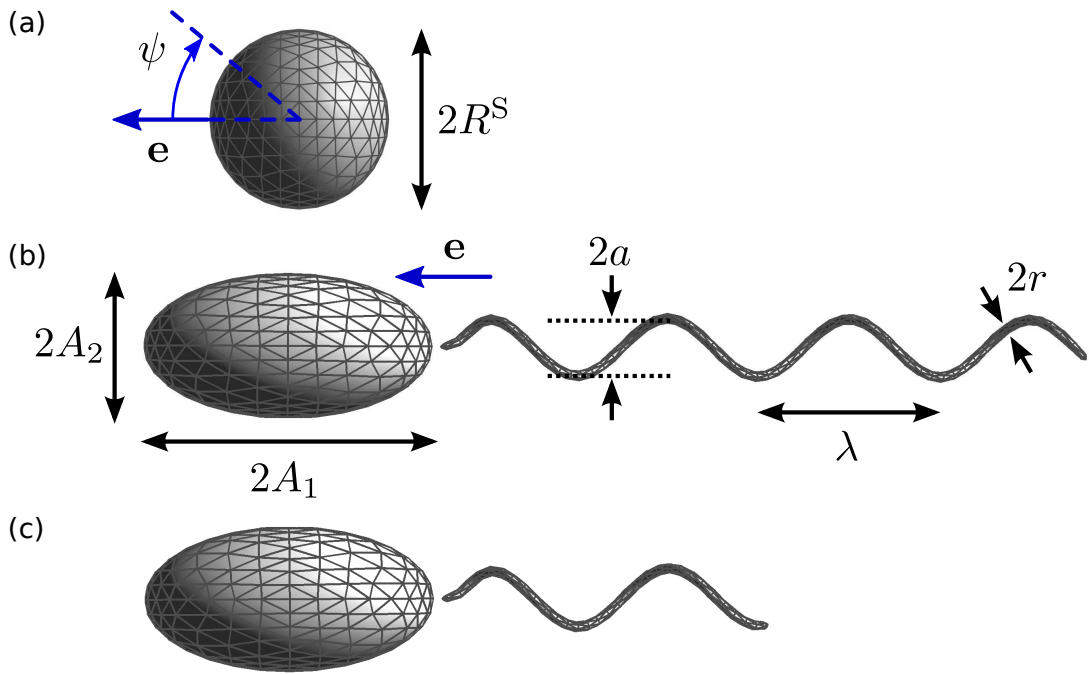}
\caption{The three model swimmers used in this numerical study. (a) The spherical squirmer, with radius $R\supS=1$. (b) The flagellated bacterium LT (long tail) with flagellum length along the centerline $L=10$, $A_1/A_2=2$, $(A_1A_2^2)^{1/3} = R\supS =1$, $\lambda=2$, $a=0.32$, and $r=0.05$. Additional modeling details are given in Shum et al.~\cite{shum_modelling_2010}. (c) Bacterium ST (short tail), with flagellum length $L=5$ and other parameters identical to those for LT.}
\label{fig:model}
 \end{figure}

For the squirmer model, the swimmer is taken to be a sphere of radius $R\supS=1$ [Fig.~\ref{fig:model}(a)]. Here and henceforth, variables are non-dimensionalized with the lengthscale set by the swimmer radius. Using only the first two modes of tangential motion in Lighthill's general squirmer model~\cite{lighthill_squirming_1952, ishikawa_hydrodynamic_2006}, we prescribe a steady surface velocity in the polar direction $\vec{e}_\psi$, 
$$
u_\psi= B_1\sin\psi + B_2\sin\psi\cos\psi,
$$
where $\psi$ is the polar angle from the swimmer's axial orientation $\vec{e}$, $B_1$ sets the swimming speed and quadrupole strength of the squirmer and $B_2$ sets the force dipole strength~\cite{ishikawa_coherent_2008, pushkin_fluid_2013}. We consider the fixed parameters $B_1=-B_2=1.5$, giving a free space swimming speed $V=(2/3)B_1=1$ and a ratio of dipole to quadrupole strengths $\beta=B_2/B_1=-1$. The negative sign of $\beta$ indicates that the swimmer is a {\em pusher}. Defining the reference point $\vec{x}\supS$ of the squirmer as its center, the fluid velocity at a point $\vec{x}$ on the surface of the squirmer is 
\begin{equation}
\vec{u}(\vec{x}) = \vec{U}\supS + \vec{\Omega}\supS\times(\vec{x}-\vec{x}\supS) + u_\psi\vec{e}_\psi(\vec{x}),
\label{eqn:squirmer_bc}
\end{equation}
where $\vec{U}\supS$ and $\vec{\Omega}\supS$ are respectively the translational and rotational velocities of the swimmer. 

The first of the model bacteria, which we shall refer to as bacterium LT (``long tail''), is shown in Fig.~\ref{fig:model}(b). It comprises a prolate spheroidal cell body and rigid, helical flagellum of finite thickness.  The semi major and semi minor axes of the cell body are $A_1=1.59$ and $A_2=0.79$, respectively, giving an aspect ratio $A_1/A_2=2$ and an effective radius $R\supS = (A_1A_2^2)^{1/3}=1$ in non-dimensionalized units. The flagellum has helical pitch $\lambda=2$, amplitude $a=0.32$, curvilinear length $L=10$ and radius $r=0.05$. It rotates with a fixed angular velocity $\omega=2\pi$ about its axis, which coincides with the major axis of the cell body $\vec{e}$.
We shall also consider bacterium ST (``short tail''), which differs only in having a shorter flagellum, $L=5$, as depicted in Fig.~\ref{fig:model}(c). 

We define the reference point $\vec{x}\supS$ in the bacterium models to be the pole of the spheroidal cell body nearer the flagellum. Imposing a no-slip condition on the cell body and flagellum, the fluid velocity at a point $\vec{x}$ on the surface of the bacterium is 
\begin{equation}
\vec{u}(\vec{x}) = \begin{cases}
               \vec{U}\supS + \vec{\Omega}\supS\times(\vec{x}-\vec{x}\supS), \qquad &\vec{x} \textrm{ on cell body,} \\
               \vec{U}\supS + (\vec{\Omega}\supS+\omega^\mathrm{M}\vec{e}) \times(\vec{x}-\vec{x}\supS),  \qquad &\vec{x} \textrm{ on flagellum.}
\end{cases}
\label{eqn:bacterium_bc}
\end{equation}

Based on previous simulations with similar parameter values~\cite{shum_hydrodynamic_2015}, and a systematic study of the effects of varying these parameters on swimmer behavior near plane boundaries~\cite{shum_modelling_2010}, we expect bacterium LT to have a tendency to swim close to surfaces and bacterium ST to swim away from surfaces~\cite{shum_modelling_2010}. Our squirmer, with parameters given above, is also expected to swim away from planar surfaces~\cite{ishimoto_squirmer_2013, spagnolie_hydrodynamics_2012}. We anticipate that the characteristic behavior near single plane boundaries will be related to the behavior near large particles, which have small surface curvatures.

In our study, the swimmer interacts with a passive spherical particle, which is modelled analogously to a squirmer except that no prescribed surface velocity $u_\psi$ is applied. The radius of the particle is $R\supP$ and the position, translational velocity, and rotational velocity vectors are respectively denoted by $\vec{U}\supP$, $\vec{\Omega}\supP$, and $\vec{x}\supP$.

A short range repulsive force between the swimmer and particle is introduced to prevent near-contact, which could lead to significant numerical errors. The repulsion acts between the points of each object nearest to one another, $\vec{x}\supS_\mathrm{near}$ and $\vec{x}\supP_\mathrm{near}$, and is of the form~\cite{takamura_microrheology_1981, ishikawa_interaction_2006}
\begin{equation}
\vec{F}^\mathrm{rep}=\alpha_1\frac{\alpha_2\exp(-\alpha_2d)}{1-\exp(-\alpha_2d)}\left(\frac{\vec{d}}{d}\right),
\label{eqn:repulsion}
\end{equation}
where $\vec{d}=\vec{x}\supS_\mathrm{near}-\vec{x}\supP_\mathrm{near}$ and $d=||\vec{d}||$.  We used the force parameters $\alpha_1=10^4$ and $\alpha_2=250$ and found that the minimum separations attained in simulations were in the range 0.04--0.06 for head-on collisions. 
Experimentally, short range forces between cells and surfaces have been measured at distances below $\sim$100\,nm~\cite{razatos_force_2000, klein_direct_2003}. Potential interactions include van der Waals forces, electrostatic interactions, and steric repulsion. We have verified that varying the range of the repulsive interaction that we used does not qualitatively affect our results. Hence we expect out findings to be broadly applicable to experimental systems provided that short range net attraction is negligible.

 
Using a boundary element method~\cite{shum_modelling_2010}, the translational and rotational velocities of the swimmer and particle are determined simultaneously by solving the equations of Stokes flow subject to the boundary velocities described above and the constraints that the hydrodynamic forces and torques on the swimmer and particle balance the forces and torques due to short range repulsion. A trapezoidal rule is used for time-stepping to construct trajectories for the swimmer and particle~\cite{shum_effects_2012}.

\section{Scattering Geometry}
\label{sec:scattering}

\begin{figure}[h!]
 \centering
  \includegraphics{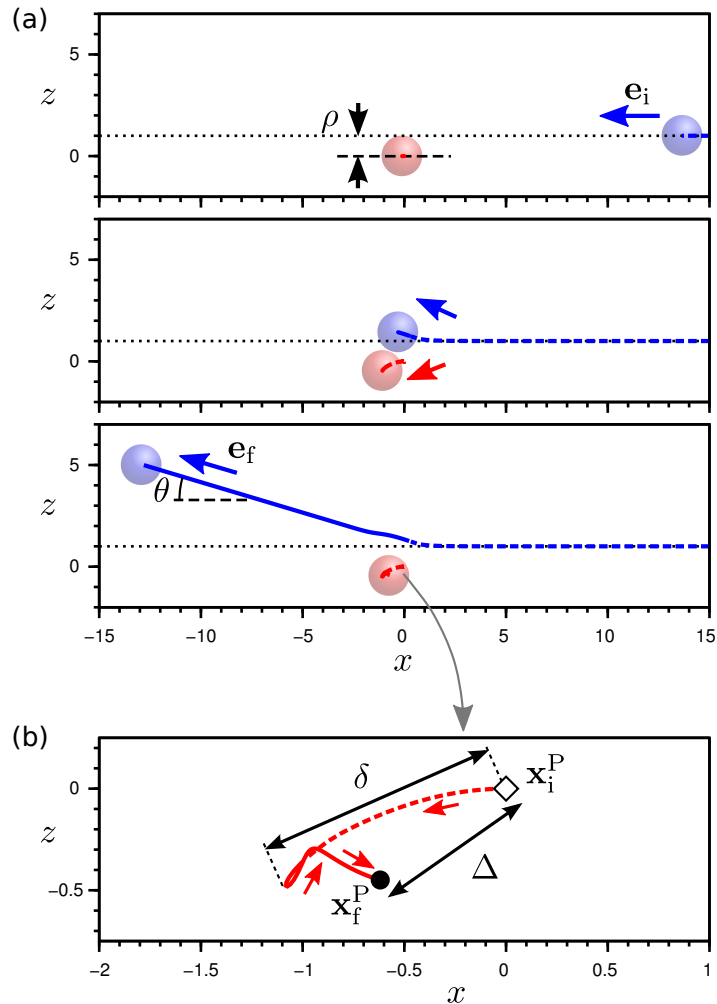}
\caption{Scattering of a squirmer by a passive particle. (a) A sequence of snapshots showing the squirmer (starting on the right, blue) approach, scatter, and swim away from a passive particle (starting on the left, red). Portions of trajectories before and after the point of closest separation between the swimmer and particle are represented by dashed and solid curves, respectively. (b) A magnified view of the trajectory of the passive particle with the displacement $\Delta$ and range $\delta$ labeled. The starting position is marked by an open diamond symbol and the final position is marked by a filled circle. Trajectories shown are simulated results for particle radius $R\supP=1$ and impact parameter $\rho=1$.}
\label{fig:scatter}
 \end{figure}

The scattering process is illustrated in Fig.~\ref{fig:scatter}, which shows the interaction between a passive particle and a squirmer. The swimmer moves in the negative $x$-direction and approaches the particle at an impact parameter $\rho$, measured from the center of the particle to the unperturbed path of the swimmer. The particle is initially centered at the origin and the initial $x$-coordinate of the swimmer is $x_\mathrm{i}=300$ (in units of $R\supS$). The final time of simulation is set such that the swimmer would have travelled a distance of 600 in the absence of the particle. Figure~\ref{fig:scatter}(a) shows three snapshots, before, during, and after close contact with the particle respectively. In this example, the particle radius is $R\supP=1$ and the impact parameter is $\rho=1$. Note that the first and last snapshots do not correspond to the initial and final times, but the squirmer orientation hardly changes when the particle is far away, so the initial and final orientation vectors $\vec{e}_\mathrm{i}$ and $\vec{e}_\mathrm{f}$ are well represented in these figures. 

For the model bacteria, the orientation and velocity vectors oscillate as the flagellum rotates relative to the cell body. The axis of the cell body precesses with a small angle around the direction of net propulsion~\cite{keller_swimming_1976}. It is therefore necessary to first compute a trajectory in the absence of the particle and then use this trajectory to determine the correct initial position and orientation of the swimmer so that, averaged over the periodic motion, the bacterium follows the intended swimming path. 

In general, the presence of the particle alters the path of the swimmer. The scattering angle $\theta$ is defined as the angle between the initial and final swimmer orientation vectors $\theta = \arccos(\vec{e}_\mathrm{i}\cdot\vec{e}_\mathrm{f})$. For bacteria, the initial and final orientations are computed as normalized averages over several periods of motion at the beginning and at the end of the trajectory respectively. Due to the rotational motion of the flagellum, the scattering interaction generally results in the bacterium and particle moving into the third dimension whereas for the squirmer, the swimmer and particle trajectories remain in the plane.

To describe the motion of the particle due to a passing swimmer, we define two characteristic quantities of the particle's path [see Fig.~\ref{fig:scatter}(b)]. The displacement $\Delta$ is the net distance moved by the particle from its initial position $\vec{x}\supP_\mathrm{i}$ to its final position $\vec{x}\supP_\mathrm{f}$. We also compute the particle's range $\delta$, defined to be the largest distance between any two points on the trajectory of the particle. We shall show that two distinct regimes of particle--swimmer interactions can be identified depending on the impact parameter. Defining the collision radius to be sum of the particle and swimmer radii, $R^\mathrm{coll}=R\supP+R\supS$, interactions are expected to be more significant when the impact parameter is smaller than the collision radius than when it is greater. We refer to these regimes respectively as {\em close approach} ($\rho<R^\mathrm{coll}$) and {\em distant passing} ($\rho>R^\mathrm{coll}$).

\section{Results}

\subsection{Analysis of individual scattering events}
\begin{figure}[h!]
 \centering
 \includegraphics[trim = 0 0 0 0, clip, width=\linewidth]{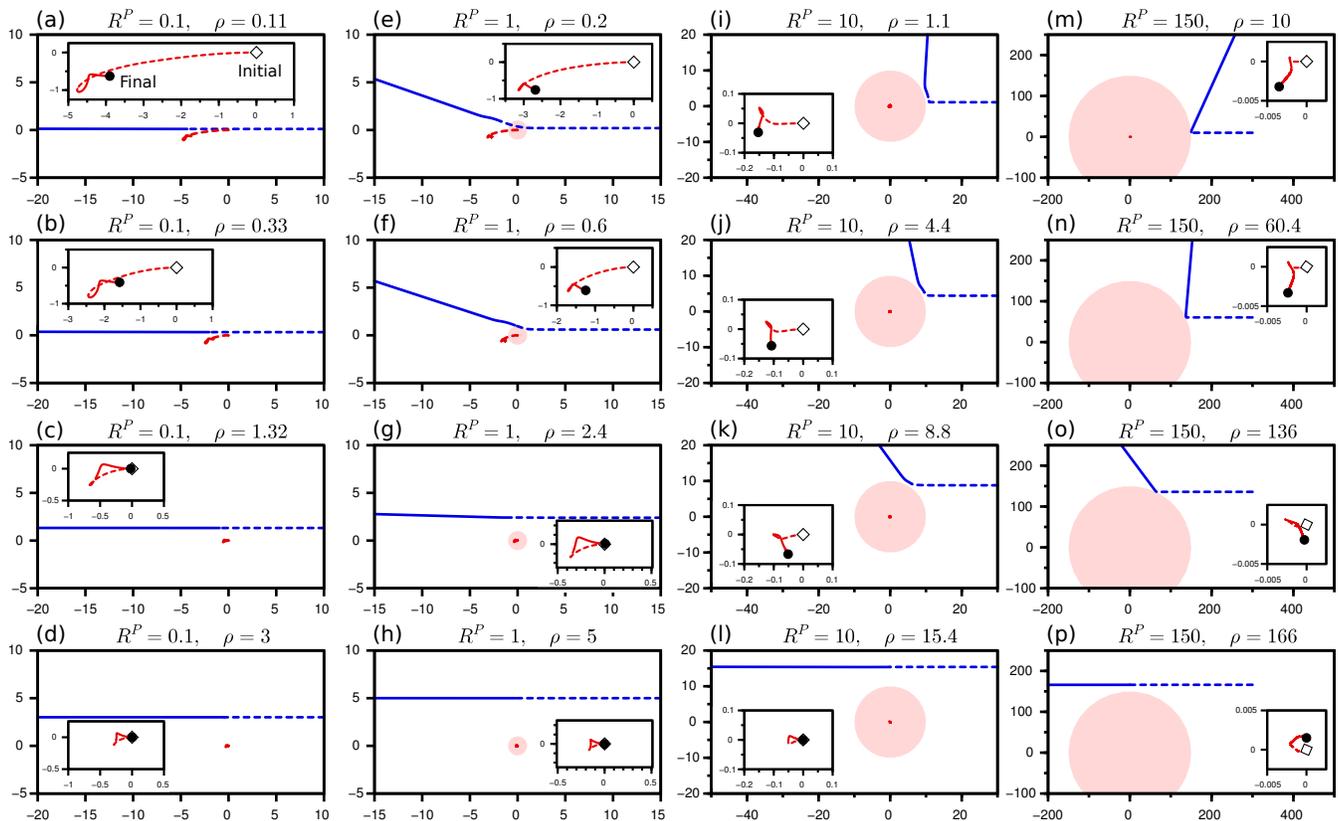}
  \caption{Swimmer and particle trajectories for a squirmer for various particle sizes and impact parameters. The swimmer (blue curves) approaches from the right. In each plot, the size of the particle is represented by a filled (red) circle centered at the origin and the trajectory of the particle center is shown as a red curve. Insets are magnifications of the particle trajectories, with starting points marked by open diamonds and final points marked by filled circles. Values of $\rho$ were selected to illustrate a progression in particle behavior from collisions near the center of the particle (top row, $\rho/R^\mathrm{coll}\leq 0.1$) to the distant passing regime (bottom row, $\rho/R^\mathrm{coll}>1$).}
\label{fig:squirmer_traj}
 \end{figure}

We first present results for the squirmer. Figure~\ref{fig:squirmer_traj} shows the particle trajectories in red and the squirmer trajectories in blue for different combinations of particle radius and impact parameter. As expected, the particles move less far when the impact parameter is larger. In addition, trajectories are almost closed loops in the distant passing regime. This is the expected behavior when the swimmer moves along an infinitely long, straight line~\cite{pushkin_fluid_2013}. For small impact parameters, small particles are pushed ahead of the approaching squirmer (to the left in our setup) for a considerable distance [almost 5 dimensionless units for $R\supP=0.1$, Fig.~\ref{fig:squirmer_traj}(a)]. Eventually, the particle slips around the swimmer and is pushed backward (to the right) by the flow field of the swimmer. Due to the front--back symmetry of the dipolar flow field, the directions of motion of the particle at the beginning and at the end of its trajectory, when it is respectively far in front of and far behind the swimmer, are approximately antiparallel. When the particle is comparable in size or larger than the swimmer [e.g., $R\supP\geq{1}$ in Fig.~\ref{fig:squirmer_traj}(e),(i),(m)], then the swimmer path is noticeably deflected by the particle at small impact parameters. The incoming direction of motion of the swimmer is not the same as the outgoing direction, hence, the directions of motion of the particle before and after collision are not antiparallel. Since the particle avoids ``backtracking,'' the initial and final particle positions maximize the distance between points on the trajectory, i.e., $\delta=\Delta$ when the particle is large and the impact parameter is small [Fig.~\ref{fig:squirmer_traj}(m),(n) insets].

\begin{figure}[h!]
 \centering
 \includegraphics[trim = 0 0 0 0, clip, width=\linewidth]{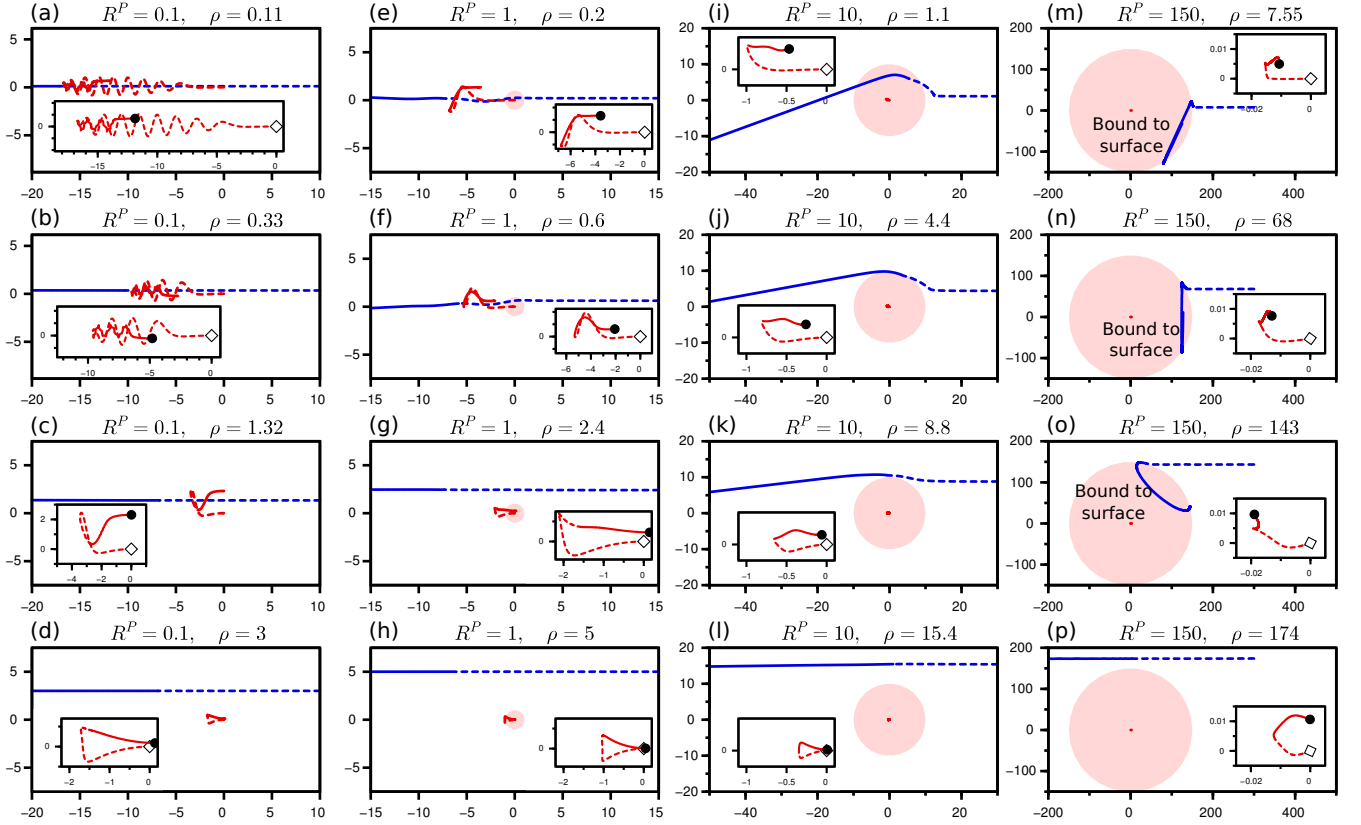}
    \caption{Swimmer and particle trajectories for bacterium LT. Insets show magnifications of particle trajectories, with starting points marked by open diamonds and final points marked by filled circles.}
  \label{fig:bacterium_A_traj}
 \end{figure}

\begin{figure}[h!]
 \centering
 \includegraphics[trim = 0 0 0 0, clip, width=\linewidth]{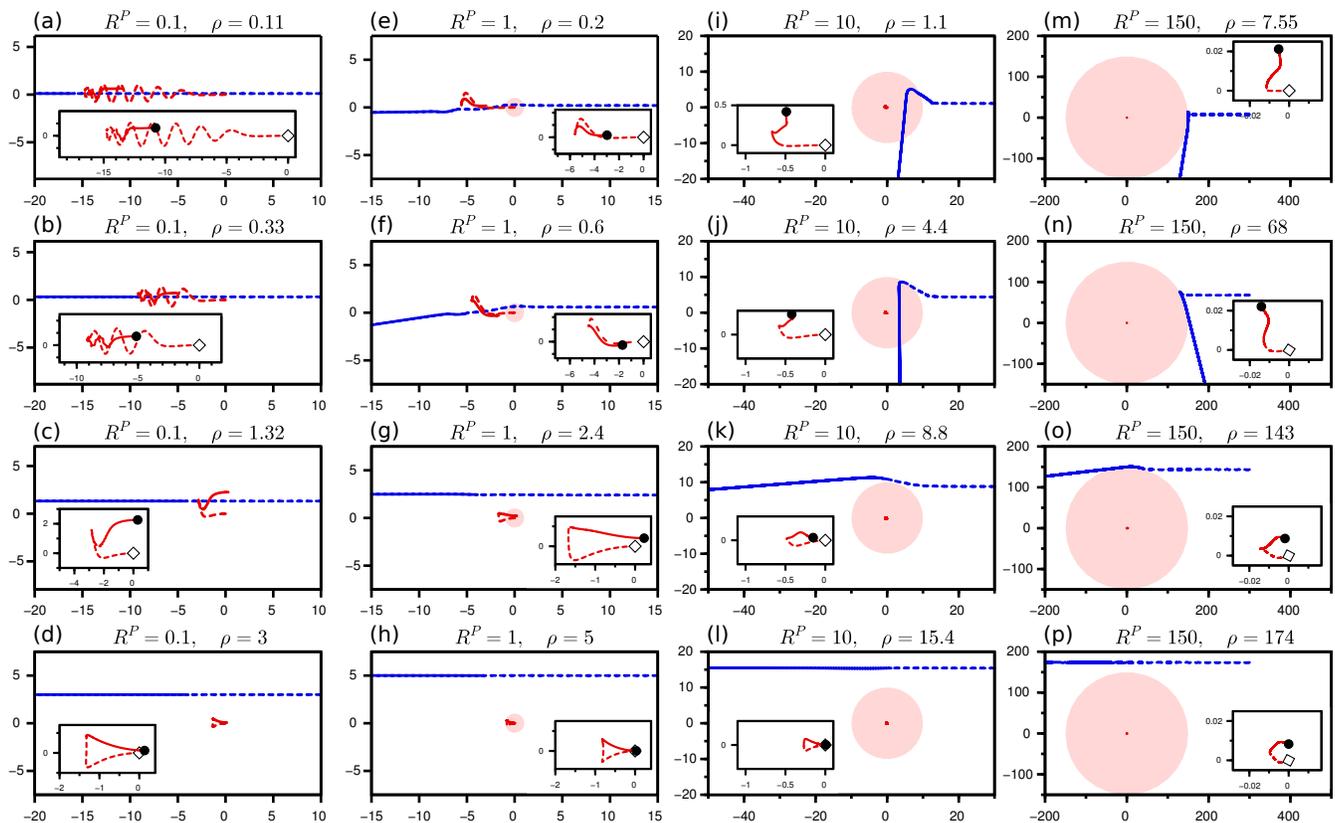}
    \caption{Swimmer and particle trajectories for bacterium ST. Insets show magnifications of particle trajectories, with starting points marked by open diamonds and final points marked by filled circles.}
  \label{fig:bacterium_B_traj}
 \end{figure}

Similar representations of bacterium--particle scattering, in Fig.~\ref{fig:bacterium_A_traj} and Fig.~\ref{fig:bacterium_B_traj}, show the same qualitative trends at large impact parameters. Trajectories differ qualitatively from those with the squirmer when the impact parameter is small, however. In the case of small particle radius ($R\supP=0.1$), the particle becomes entrained in the rotational near field flow around the bacterium, resulting in helical portions in the trajectory of the particle [Fig.~\ref{fig:bacterium_A_traj}(a) and Fig.~\ref{fig:bacterium_B_traj}(a)]. The differences between near field fluid flows of the squirmer and bacterial swimmer also lead to differences in the deflection of the swimmers. The squirmer is always deflected away from the particle, akin to elastic collisions [e.g., Fig.~\ref{fig:squirmer_traj}(j)]. The bacteria, however, tend to be deflected in the opposite direction so as to follow the curvature of the particle [e.g., Fig.~\ref{fig:bacterium_A_traj}(j) and Fig.~\ref{fig:bacterium_B_traj}(k)] and are also deflected in the out-of-plane direction.
  
Qualitatively, the two bacterium models show the same behavior in particle motion and swimmer deflection except when the particle is very large ($R\supP=150$). In this case, bacterium LT becomes hydrodynamically bound to the surface of the particle if the impact parameter is sufficiently small ($\rho\leq{}R^\mathrm{coll}=R\supP+R\supS$). The swimmer traces out circular orbits close to the surface of the particle [Fig.~\ref{fig:bacterium_A_traj}(m)--(o)]. This mirrors the circular motion expected for bacterium LT near a plane wall. Bacterium ST swims away from plane walls and, therefore, is not expected to become hydrodynamically bound to a spherical particle of any radius.

\begin{figure}[h!]
 \centering
 \includegraphics[trim = 0 0 0 0, clip, height=18cm]{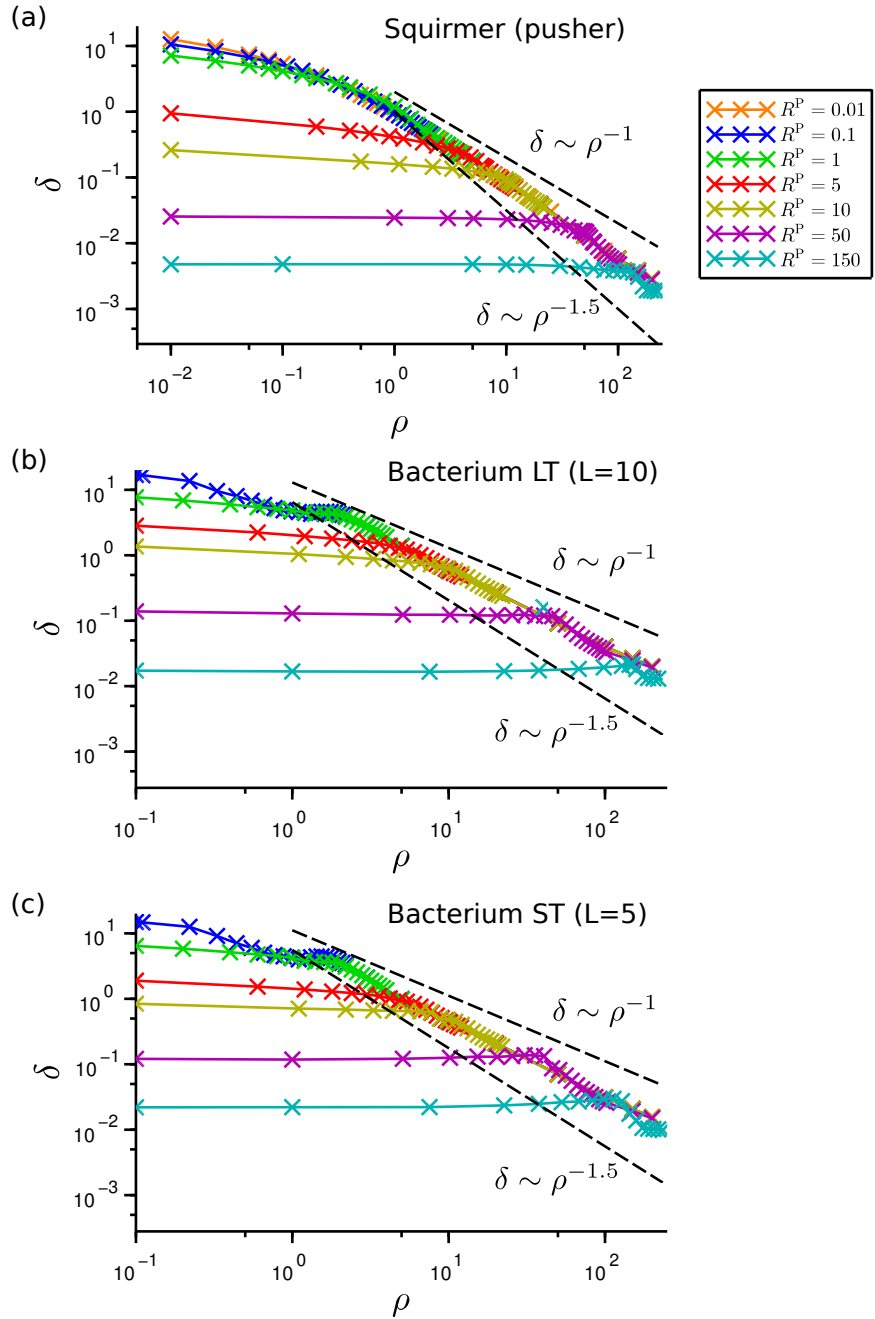}
\caption{Particle range as functions of impact parameter for (a) the squirmer, (b) bacterium LT, and (c) bacterium ST. Colors correspond to particle sizes indicated in the legend in (a). The case $R\supP=0.01$ is only shown for the squirmer.}
\label{fig:scaling}
 \end{figure}

The range of the particle's motion $\delta$ is plotted as a function of particle radius and impact parameter for all three swimmer models in Fig.~\ref{fig:scaling}. We note that there is no significant qualitative difference among the three swimmers. The range is greatest when the impact parameter is small and when the particle radius is small. The curves for different particle sizes coincide for impact parameters large compared with $R^\mathrm{coll}$, indicating that large and small particles are transported equally well by distantly passing swimmers. Figure~\ref{fig:scaling} shows that in this large impact parameter regime, the particle range approximately fits a power law scaling $\delta\sim\rho^{-\alpha}$ with exponent $1<\alpha<1.5$.

To understand the origin of this scaling, we analyzed the exact solutions to a far field approximation of the scattering problem (see Appendix~A). From this analysis, we expect the particle range to scale as $\delta\sim\rho^{-1}$ at small impact parameters and as $\delta\sim\rho^{-2}$ at large impact parameters. The cross-over takes place around $\rho\sim x_\mathrm{i}=300$, which is slightly above the maximum impact parameter used in full simulations of swimmers with particles. Thus, the dependence of $\delta$ on $\rho$ shown in Fig.~\ref{fig:scaling} (in the distant passing regime) exemplifies the intermediate $\rho$ behavior, before the transition to $\delta\sim\rho^{-2}$ scaling.

\begin{figure}[h!]
 \centering
 \includegraphics[trim = 0 0 0 0, clip, height=18cm]{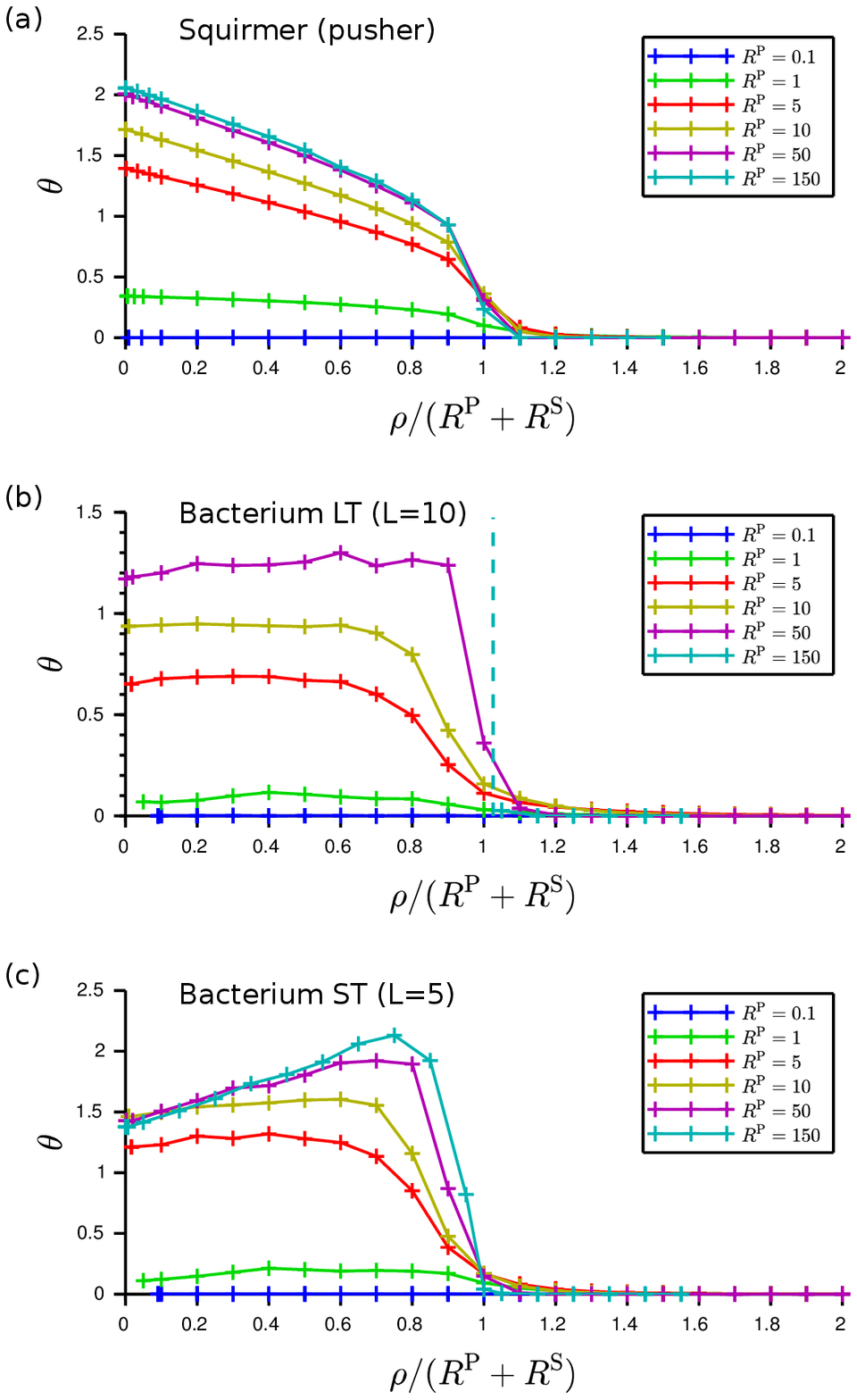}
\caption{Deflection angles as functions of impact parameter rescaled by collision radius for (a) the squirmer, (b) bacterium LT, and (c) bacterium ST.}
\label{fig:angle}
 \end{figure}

It is apparent from Figs.~\ref{fig:squirmer_traj}, \ref{fig:bacterium_A_traj} and \ref{fig:bacterium_B_traj} that the path of the swimmer is not significantly altered by the small particles or when the impact parameter is large. Conversely, large particles noticeably deflect swimmers when the impact parameter is small. Moreover, the deflection depends sensitively on the details of the swimmer model. To quantify this we plot the deflection angles of the three swimmers as functions of $\rho / R^\mathrm{coll}$, the impact parameter scaled by the collision radius of the swimmer with the particle, in Fig.~\ref{fig:angle}. For all swimmers, the deflection angle rapidly decreases to zero as the impact parameter increases beyond the collision radius. Hence, swimmers must approach the particle closely for deflection to occur. In general, larger particles cause greater deflection but the effect appears to saturate as there was little difference between the deflection curves (as functions of the normalized impact parameter $\rho/R^\mathrm{coll}$) for the two largest particles in the cases of the squirmer and bacterium ST. Deflection angles are as large as 2 radians in some instances.
 
For squirmers, the deflection angle is greatest as $\rho\rightarrow 0$ and decreases gradually with $\rho$ until $\rho\approx{}R^\mathrm{coll}$. For bacterium LT, deflection angles were found to be almost independent of impact parameter for most of the close approach regime. As discussed above, bacterium LT is also unique in that the swimmer becomes hydrodynamically bound to the largest particle. The deflection angle is therefore not well defined for impact parameters below the critical trapping value, indicated approximately by the vertical dashed line in Fig.~\ref{fig:angle}(b). The deflection of bacterium ST exhibits the opposite trend to the squirmer; the deflection angle increases gradually with impact parameter (below $R^\mathrm{coll}$) and, for large particles, peaks at a value close to the collision radius.

To test the sensitivity of our results to the choice of parameters for short range repulsion [Eq.~(\ref{eqn:repulsion})], we varied $\alpha_2$ while keeping $\alpha_1\alpha_2$ constant for the case of a squirmer interacting with a colloid of radius $R\supP=1$. No qualitative differences in behavior were noted. With $\alpha_2=50$, which produces the same repulsive force as the baseline $\alpha_2=250$ case at a separation five times larger, the resulting particle range $\delta$ was 6\,\% larger averaged over the impact parameters that necessitate repulsive interactions. With $\alpha_2=25$, the particle range was an average of 23\,\% larger than with $\alpha_2=250$.

\subsection{Effective swimmer diffusivity}

The dependence of scattering angle on impact parameter shown in Fig.~\ref{fig:angle} can be used to calculate an effective diffusivity for swimmers in colloid suspensions, assuming that swimmers are dilute so that interactions among them may be neglected. Since our simulations show that swimmer paths are straight except when the swimmer is almost in contact with a particle, it is reasonable to treat scattering in suspensions as a series of events with isolated particles provided that there is rarely more than one particle within a distance of $\sim{}R^\mathrm{coll}$ from the swimmer at any given time.

Berg~\cite{berg_random_1993} gave the effective diffusion coefficient $D\supS$ of a tumbling, self-propelled swimmer travelling with constant speed $v$ and mean run duration $\tau$ as
\begin{equation}
D\supS = \frac{v^2\tau}{3(1-\alpha)},
\label{eqn:swimmer_diff_berg}
\end{equation}
where $\alpha$ is the mean cosine of the angle between swimming paths before and after a tumble. It is assumed that the swimmer follows a three-dimensional path of straight line segments connected by random, instantaneous direction changes. Applying this model to reorientations due to collisions with colloids, the run duration can be estimated as $\tau = [n\supP{}v\pi (R^\mathrm{coll})^2]^{-1}$, where $n\supP$ is the number density of particles in the suspension, and $\alpha$ can be determined from the scattering angles shown in Fig.~\ref{fig:angle} using the formula,
\begin{equation}
\alpha = \frac{2}{(R^\mathrm{coll})^2}\int_0^{R^\mathrm{coll}}\rho \cos[\theta(\rho)]\dd{\rho} = 2\int_0^1\rho' \cos[\theta(\rho')]\dd{\rho'},
\label{eqn:alpha}
\end{equation}
where $\rho'=\rho/R^\mathrm{coll}$.

Note that this definition neglects scattering that takes place when $\rho>R^\mathrm{coll}$ but this is not expected to significantly affect the calculated diffusivity since the scattering angle is small at larger impact parameters. Defining a larger collision radius to take these events into account would decrease the average scattering angle but increase the overall frequency of collisions. 

We remark that by symmetry, all scattering events between a particle and our spherical squirmer remain planar. In a suspension of randomly distributed particles, however, we expect each pairwise interaction to lie in different scattering planes. Hence, the overall path of the squirmer would not be planar but three-dimensional. For bacteria, each scattering event is three-dimensional.
  
To facilitate comparisons between suspensions of colloids of different sizes, we define the particle volume fraction $\phi\supP=4\pi n\supP (R\supP)^3/3$ and obtain an expression for the swimmer diffusivity,
 \begin{equation}
D\supS = \frac{4v(R\supP)^3}{9\phi\supP(R^\mathrm{coll})^2(1-\alpha)}.
\label{eqn:swimmer_diff}
\end{equation}
From this expression, we predict that the diffusive spreading of swimmers in a suspension is monotonically reduced as the colloid volume fraction increases. The explanation for this is that the swimmer encounters obstacles and changes direction more frequently, hindering its net displacement. There is also a trivial dependence on the swimming speed $v$, which sets the only time scale in this problem. Hence, we define a normalized effective swimmer diffusivity,
\begin{equation}
\tilde{D}\supS = D\supS\phi\supP/(R\supS v),
\end{equation}
where the factor $R\supS=1$ is superfluous in dimensionless units but yields a consistent, dimensionless value for $\tilde{D}\supS$ when variables on the right hand side are replaced by their dimensional counterparts.

\begin{figure}[h!]
 \centering
 \includegraphics{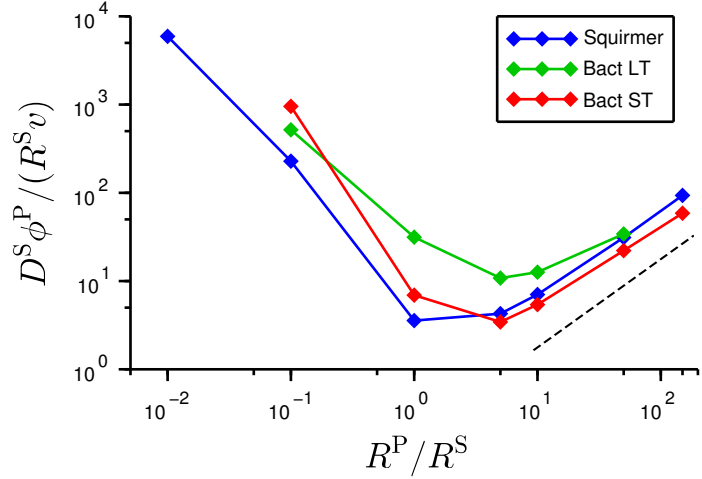}
\caption{Effective diffusivities of a swimmer in a random suspension of colloids as functions of the relative size of colloids. The swimmer is assumed to move in straight lines except when scattered by colloids. The dashed line is a guide to the eye showing the relationship $D\supS\sim R\supP v/\phi\supP$.}
\label{fig:swimmer_diff}
 \end{figure}

The normalized effective diffusivities of our three swimmers are plotted in Fig.~\ref{fig:swimmer_diff} as functions of the particle radius. For all swimmers, the diffusivity is minimized at an intermediate value of $R\supP$ comparable to the swimmer size. Small particles cause insignificant deflection of the swimmer, yielding $\alpha\approx 1$, which leads to large values of the effective diffusivity. 

In the limit of large particles, Fig.~\ref{fig:scatter} showed that scattering angles become functions of $\rho/R^\mathrm{coll}$ that are relatively insensitive to particle size (excluding bacterium LT, which becomes hydrodynamically bound to large particles). We can explain this saturation in deflection by considering a swimmer interacting with a fixed no-slip plane, tangential to the particle surface at the point of collision. Assuming that the swimmer is negligibly deflected before collision, the orientation of the tangent plane depends only on the ratio $\rho/R\supP\approx \rho/R^\mathrm{coll}$. Since the direction of the swimmer after scattering depends only on the orientation of the plane, the generic behavior for a boundary escaping swimmer interacting with a large particle is that $\theta=\theta(\rho/R^\mathrm{coll})$.

Consequently, the mean cosine of the scattering angle, given by Eq.~\ref{eqn:alpha}, is insensitive to $R\supP$. Further noting that $R\supP\approx R^\mathrm{coll}$ for large particles, Eq.~\ref{eqn:swimmer_diff} reduces to $D\supS \sim v^2\tau \sim R\supP v/\phi\supP$. This scaling is evident for $R\supP/R\supS>10$ in Fig.~\ref{fig:swimmer_diff}. 

Our results show that swimmer diffusivity becomes large when the particle size becomes either very small or very large. As discussed by Berg~\cite{berg_random_1993}, however, the persistence length in real swimmers will be limited by Brownian rotation. The assumption that swimmers travel along straight paths between collisions breaks down when reorientation due to collisions is insignificant, i.e., when the volume fraction is low or when particles are very small or very large compared with the swimmer. Corrections to Eq.~\ref{eqn:swimmer_diff_berg} necessary to account for Brownian rotation and swimmer tumbling can be made as discussed by Lovely and Dahlquist~\cite{lovely_statistical_1975}. In the present work, we focus on the role of scattering and neglect other sources of reorientation.

Another assumption in the derivation of Eq.~\ref{eqn:swimmer_diff_berg} is that reorientations occur instantaneously. In our simulations, we found that swimmer reorientation was only significant when the particle was large and almost in contact with the swimmer. Except in cases where the swimmer became hydrodynamically bound to the particle, the period of near contact lasted for less than 15\% of the simulation time (for a swimmer path length of $600R\supS$). Hence, we expect the overall fraction of time spent turning to be negligible in a dilute suspension.

\subsection{Effective particle diffusivity}
\label{sec:particle_diff}
We follow the approach of Thiffeault and Childress~\cite{thiffeault_stirring_2010} to estimate the effective diffusivity of particles in a dilute, isotropic bath of swimmers travelling along infinite paths. In this analysis, we consider a single, spherical particle in a bath of swimmers and neglect interactions among swimmers. We also neglect thermal Brownian motion of the particle; while Brownian motion adds to the overall diffusivity of the particle, it has been suggested that entrainment by swimmers at small impact parameters could be reduced by these fluctuations~\cite{mathijssen_contact_nodate}. The relative significance of thermal effects can be minimized under certain conditions, such as in high viscosity fluids, when particles are large, or when swimmers are fast.

The mean squared displacement of particles over a long time period $t$ is given by
\begin{equation}
<||\vec{x}\supP{}(t)-\vec{x}\supP{}(0)||^2>=6D\supP t=2\pi v n\supS t \int_0^\infty \rho\Delta^2(\rho)\dd{\rho},
\label{eqn:msd}
\end{equation}
where $n\supS$ is the average number of swimmers per unit volume and $\Delta$ is the net displacement of the particle due to a single swimmer approaching with impact parameter $\rho$. Our simulation data for $\Delta(\rho)$ is based on finite, but long (path length $\sim 600$), swimmer trajectories. Hence, our calculations neglect the small displacements that would occur if the swimmer paths were infinitely extended forward and backward in time. More accurate formulas for the mean squared displacement due to swimmers with finite path lengths have been proposed~\cite{lin_stirring_2011, pushkin_fluid_2013} but these require integration over two- or three-dimensional impact parameter space, which is time consuming when the coupled trajectories of the swimmer and particle are computed to high accuracy.

\begin{figure}[h!]
 \centering
 \includegraphics[trim = 0 0 0 0, clip, width=\linewidth]{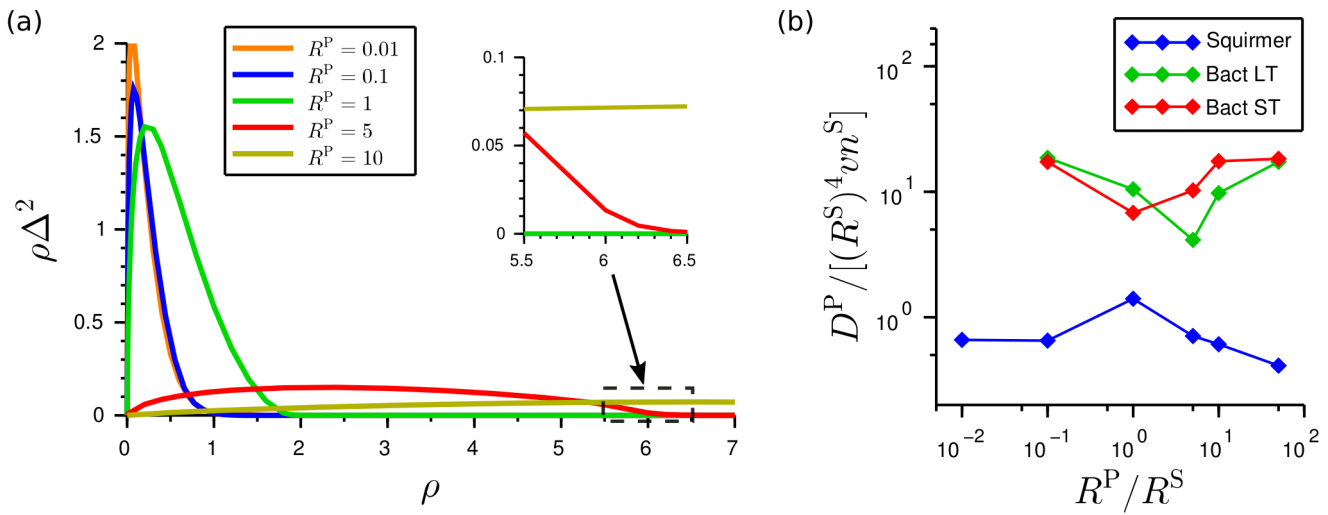}
\caption{Effective diffusivity of particles due to passing swimmers. (a) Contributions to diffusive particle motion due to a squirmer approaching at different impact parameters. The function $I(\rho)=\rho\Delta^2(\rho)$ represents the squared displacement due to a collision at impact parameter $\rho$ weighted by the probability of occurrence of this impact parameter ($\sim \rho$). (b) Particle diffusivities normalized by $(R\supS)^4vn\supS$ for three swimmer models as functions of relative particle radius.}
\label{fig:particle_diff}
 \end{figure}

To illustrate the effect of a particle's size on its diffusivity (neglecting thermal motion), we consider interactions with the squirmer. The integrand in Eq.~(\ref{eqn:msd}), $I(\rho)=\rho\Delta^2(\rho)$, is plotted in Fig.~\ref{fig:particle_diff}(a) for different particle sizes. For each value of $R\supP$, there is a prominent peak, which shifts to larger values of $\rho$ and becomes lower and broader as the particle size increases. In each case, the integrand is small when $\rho>R^\mathrm{coll}$ and, in fact, the integrands for different particle sizes overlap in the distant passing regime. This mirrors the observed overlap in curves of $\delta$ at large impact parameters (Fig.~\ref{fig:scaling}).

At small impact parameters, $I$ is generally greater for smaller particles than for larger particles, i.e., small particles are entrained less than large particles. This trend can also be seen from trajectories in Figs.~\ref{fig:squirmer_traj}--\ref{fig:bacterium_B_traj} and plots of the particle range in Fig.~\ref{fig:scaling}. Interestingly, however, Fig.~\ref{fig:particle_diff}(a) shows that for any pair of particle sizes, there is a range of impact parameters for which the smaller particle experiences less net displacement $\Delta$ than the larger particle at the same $\rho$. One might hypothesize that this is simply due to the larger particle occupying a larger volume and, hence, being closer to the swimmer in terms of the distance between the surface of the particle and the swimmer. While it is true that the hydrodynamic stress distribution on the larger particle is different from that on the small particle, we find no significant difference in the resulting displacement of the two particles when the impact parameter is only slightly larger than the collision radius of the larger particle. Comparing two particles of sizes $R\supP=0.1$ and $R\supP=5$, respectively, with impact parameter $\rho=6.2$, for example, the minimum separation between a squirmer and the smaller particle is $0.3$ while the minimum separation for the larger particle is more than $5$. Despite this large difference in nearest separations, the trajectories of the particle centers are almost identical and final displacements are both small [Fig.~\ref{fig:particle_diff}(a) inset]. In addition, note that when $\rho<0.5$, the smaller particle is displaced much more than the larger one. Thus, displacement is not simply a function of surface-to-surface proximity.

Instead, we attribute the enhanced displacement of larger particles to the deflection of the swimmer upon collision. A swimmer that distantly passes a small particle is not deflected and pushes the particle forward on approach, then backward after passing by. In contrast, a particle large enough to obstruct the path of the swimmer and cause a deflection might instead be pushed forward as the swimmer approaches and sideways after collision. The final displacement of the large particle could then be greater because there is less cancellation of displacements.

We estimate the total effective diffusivity of particles in a dilute bath of uncorrelated swimmers using the formula
\begin{equation}
D\supP = \frac{\pi}{3} v n\supS \int_0^{R^\mathrm{thresh}} \rho\Delta^2(\rho)\dd{\rho},
\label{eqn:particle_diff}
\end{equation}
where $R^\mathrm{thresh}$ is a cut off distance larger than the collision radius. We set $R^\mathrm{thresh}=60$ and compute $D\supP/(v n\supS)$ for particles up to $R\supP=50$. It has been experimentally demonstrated that the diffusivity of tracer particles due to a dilute suspension of swimmers scales proportionally with the active flux, defined as $J^\mathrm{A}=v n\supS$~\cite{jepson_enhanced_2013, jeanneret_entrainment_2016, mino_enhanced_2011}, as expected from Eq.~\ref{eqn:particle_diff}. This motivates our choice of normalization,  $\tilde{D}\supP=D\supP/(v n\supS)$. In terms of dimensional variables, this dimensionless characterization is written $\tilde{D}\supP=D\supP/[(R^\mathrm{S})^4vn^\mathrm{S}]$. 

As shown in Fig.~\ref{fig:particle_diff}(b), a particle's normalized diffusivity does not trivially decrease with its size. Moreover, the qualitative dependence on particle radius is not the same for all swimmer types. For the squirmer, there is a peak in diffusivity at a particle radius comparable to the swimmer radius. In contrast, the two bacterium models exhibit a minimum in particle diffusivity at an intermediate value of $R\supP$ comparable to the swimmer size. Figure~\ref{fig:particle_diff}(b) also suggests that the bacterium models result in values of $\tilde{D}\supP$ an order of magnitude greater than those from the squirmer. We note, however, that the lengthscale in our model is set by the cell body of the bacterium and does not account for the flagellum. If we were to instead consider the length of the swimmer, including the flagellum, as the characteristic swimmer size, then the calculated particle diffusivities would be comparable to those of the squirmer.

For comparison, we quote the reported (dimensional) quantity $\tilde{D}\supP (R\supS)^4=13$\micron{}$^4$ obtained experimentally for latex particles of radius $R\supP=1$\micron{} in suspensions of \textit{E.\ coli}~\cite{mino_induced_2013}. Another study, using nonmotile cells as the tracers, obtained the value $\tilde{D}\supP (R\supS)^4=7.1$\micron{}$^4$~\cite{jepson_enhanced_2013}. Assuming $R\supS=0.8$\micron{} (for an ellipsoid of major axis 2\micron{} and minor axis 1\micron{}), our theoretical analysis for the bacterial models estimates $\tilde{D}\supP (R\supS)^4\approx 4$\micron{}$^4$, which is of the same order of magnitude as experimental results.

\section{Discussion}
While the motion of low Reynolds number swimmers near plane boundaries has been extensively studied theoretically~\cite{ramia_role_1993, berke_hydrodynamic_2008, shum_modelling_2010, giacche_hydrodynamic_2010, drescher_fluid_2011, spagnolie_hydrodynamics_2012}, there have been relatively few analyses of the hydrodynamic influence of curved surfaces or free particles on swimmers. Experiments have shown that spermatozoa tend to swim along corners of microchannels~\cite{nosrati_predominance_2016} and can follow the convex curvature of a slowly bending channel~\cite{denissenko_human_2012}. Similarly, bacteria have been observed to orbit around the base of large, circular pillars protruding from a flat substrate~\cite{sipos_hydrodynamic_2015}. When the pillar radius was larger than 100\micron{}, the fraction of cells trapped at the pillar after collision was comparable to the fraction trapped by a flat wall. Since the radius of curvature of such pillars is much larger than the body length of a bacterium, the intersection between the pillar and the substrate is similar to the intersection between perpendicular planes, which was shown to be hydrodynamically attractive to some swimming model bacteria~\cite{shum_hydrodynamic_2015_b}.

Interestingly, self-propelled gold--platinum nanorods were found to similarly orbit around spherical particles resting on a substrate even when the diameter of the particle was as small as half of the length of the rod~\cite{takagi_hydrodynamic_2014}. Motivated by these experimental observations, a detailed mathematical model was proposed by Spagnolie et al.~\cite{spagnolie_geometric_2015} to describe hydrodynamic interactions between a dipolar swimmer and a spherical particle that was assumed to be stationary. Using this model, hydrodynamic trapping of the swimmer was predicted for particles larger than a critical radius, which depended on the relative dipole strength and aspect ratio of the swimmer. It is known, however, that near field effects not captured by the dipolar flow field approximation also influence hydrodynamic trapping at surfaces~\cite{spagnolie_hydrodynamics_2012}. This is evident from the non-trivial transitions in stability of near-surface swimming in squirmers when squirming modes, corresponding to the relative strengths of the dipole and higher order multipoles, are varied~\cite{ishimoto_squirmer_2013}.

Nevertheless, our current simulations demonstrate hydrodynamic trapping of swimmers at curved surfaces similar to the experimental observations for sperm~\cite{denissenko_human_2012}, bacteria~\cite{sipos_hydrodynamic_2015, jashnsaz_hydrodynamic_2017} and catalytic nanorods~\cite{takagi_hydrodynamic_2014}. In our study, entrapment only occurred for the swimmer that is hydrodynamically bound to planar boundaries (bacterium LT) and required the particle radius to be about 150 times the volumetric radius of the bacterial cell body, corresponding to a physical size of $R\supP\approx 100$\micron{} for bacteria~\cite{shum_modelling_2010}. This coincides with the critical pillar radius for trapping bacteria in the study by Sipos et al.~\cite{sipos_hydrodynamic_2015}, though we note that the estimated critical sphere radius in our model is expected to be sensitive to the geometry of the model swimmer. Spagnolie et al.~\cite{spagnolie_geometric_2015} suggested that the swimmer would need to collide with or pass very close to a sphere for hydrodynamic trapping to occur. This is supported by our simulations of bacterium LT, which passed the $R\supP=150$ particle without becoming trapped when the impact parameter was $\rho\geq 1.05 R^\mathrm{coll}$.

For swimmer--particle interactions that do not result in trapping, we find that close approaches to the particle result in deflection of the swimmer path. The deflection is insignificant for particles smaller than the swimmer. Unsurprisingly, large particles cause greater swimmer deflection, though this effect starts to saturate as particle radii increase above 10 times the swimmer length scale for bacterium ST and the squirmer (which do not become hydrodynamically bound to large particles).

The scattering of the swimmer upon collisions with particles sets an effective timescale of reorientation for an otherwise straight swimming organism in a suspension. Using simulations to understand the details of the scattering distributions, we have shown that the normalized effective diffusivity of a swimmer in a suspension of given volume fraction depends non-monotonically on the particle radius and is minimized when the swimmer and particles are comparable in size. 

Many theoretical studies have sought to quantify the influence of active particles on the motion of passive colloids in a fluid~\cite{dunkel_swimmer-tracer_2010, thiffeault_stirring_2010, lin_stirring_2011, pushkin_fluid_2013-1, pushkin_fluid_2013}. For analytical treatment, the displacement of a tracer due to the flow field of a passing swimmer is commonly obtained by integrating the time-varying flow velocity at a fixed tracer position~\cite{dunkel_swimmer-tracer_2010, pushkin_fluid_2013-1}. This approach is valid in the far field when the tracer motion is small enough that Lagrangian effects may be neglected. If a swimmer approaches close to the tracer, however, Lagrangian effects are important and lead to entrainment of the particle with the moving swimmer. Thus, it has been proposed that the effective diffusion coefficient characterizing the motion of tracers in a bath of swimmers could conceptually be expressed as the sum of three contributions~\cite{pushkin_fluid_2013},
\begin{equation}
D\supP \approx D_{rr} + D_\mathrm{entr} +D_\mathrm{therm},
\label{eqn:diffusivity}
\end{equation}
where $D_{rr}$ is due to the far field flow of uncorrelated swimmer trajectories of finite persistence lengths (infinite, straight trajectories would cause no net displacement of a tracer in the far field), $D_\mathrm{entr}$ is the near field entrainment term and $D_\mathrm{therm}$ is a contribution from thermal noise. We emphasize that Eq.~\ref{eqn:diffusivity} is not a mathematically rigorous decomposition but is simply intended to express insight into the physics on a qualitative level. 

Our results for the motion of finite size particles due to swimmers show that the particle's motion is almost independent of its radius in the distant passing regime, where the trajectory of the swimmer is unaffected by the particle. Hence, we expect $D_{rr}$ to be unchanged when particles of finite size are considered instead of tracers. In head-on collisions, small particles tend to be entrained over long distances while large particles are not pushed far before the swimmer changes direction and swims away. We interpret this as a reduction of $D_\mathrm{entr}$ with increasing particle size.

The scattering of swimmers upon collision with finite size particles also gives rise to a new contribution to particle diffusivity $D_\mathrm{scat}$ not present in the description of (point-like) tracer diffusion, Eq.~(\ref{eqn:diffusivity}). Unlike random, ``tumbling'' reorientations and gradually curving trajectories, which are intrinsic to the swimmer and accounted for in the $D_{rr}$ term~\cite{pushkin_fluid_2013}, swimmer deflections due to collisions with particles take place at specific times, namely, when the swimmer encounters the surface of a particle. The perturbation of the swimmer path results in a different particle trajectory. In some cases, the ``closed loop'' path that a tracer would follow is opened up by swimmer scattering, leading to a large net displacement and increased diffusivity of finite size particles. 

Hence, the net effect of increasing particle size, taking into account the reduction of $D_\mathrm{entr}$ and the increase of $D_\mathrm{scat}$, is a non-trivial and swimmer-specific dependence of the effective diffusivity of a particle on $R\supP$. Diffusivities are similar at the two extremes in particle size, while intermediate sizes can lead to either enhanced or diminished diffusivities.  

We note that our approaches for estimating the effective diffusivities of swimmers and particles do not take into account the potential changes in swimming speed when a particle is nearby. From Eqs.~(\ref{eqn:swimmer_diff}) and (\ref{eqn:particle_diff}), we expect both diffusivities to decrease if the average swimming speeds are lower due to the presence of particles. Previous simulation studies have suggested that bacteria swimming very close and parallel to a no-slip wall could be 10\% faster than in bulk fluid~\cite{ramia_role_1993}, though this assumes a fixed motor speed. For a given power or motor torque, the increased drag due to the presence of the nearby boundary results in a lower swimming speed for model bacteria~\cite{shum_modelling_2010}. The speed also depends on the orientation relative to the surface. It is therefore not simple to predict how the average speed of a swimmer would depend on the size or volume fraction of suspended particles. 

Experimentally, however, bacteria have been shown to swim at a nearly fixed speed under confinement in channels with heights down to 3\micron{}~\cite{biondi_random_1998} and the speeds of self-propelled nanorods are also unchanged when orbiting a spherical obstacle~\cite{takagi_hydrodynamic_2014}. In light of these observations, we expect that changes in swimming speed in suspensions or particles would also be negligible, at least for some types of swimmers.

\begin{acknowledgments}
We acknowledge funding from the ERC Advanced Grant MiCE and thank Arnold J.~T.~M.~Mathijssen for helpful discussions.
\end{acknowledgments}

\appendix

\section{Far field scaling of particle range}
\label{sec:farfieldscaling}
We seek an approximate analytical solution for the particle range $\delta$ in the limit of large impact parameters. The swimmer--particle scattering problem is as described in Sec.~\ref{sec:scattering}. For a force-free, torque-free swimmer, the far field fluid flow is given by the Stokes dipole term. For large impact parameters, swimmer deflection by the particle is negligible so we assume that the swimmer moves with constant velocity $\vec{V} = V\vec{e}$, where $\vec{e}=(-1,0,0)^\mathrm{T}$, from the initial point $\vec{x}\supS_\mathrm{i}=(x_\mathrm{i},0,\rho)^\mathrm{T}$ at time $t=0$ to the final point $\vec{x}\supS_\mathrm{f}=(-x_\mathrm{i},0,\rho)^\mathrm{T}$ at time $t=T_{\max}=2x_\mathrm{i}/V$. The particle is initially at the origin. 

\begin{figure}[h!]
	\centering
	\includegraphics[trim = 0 0 0 0, clip, width=\linewidth]{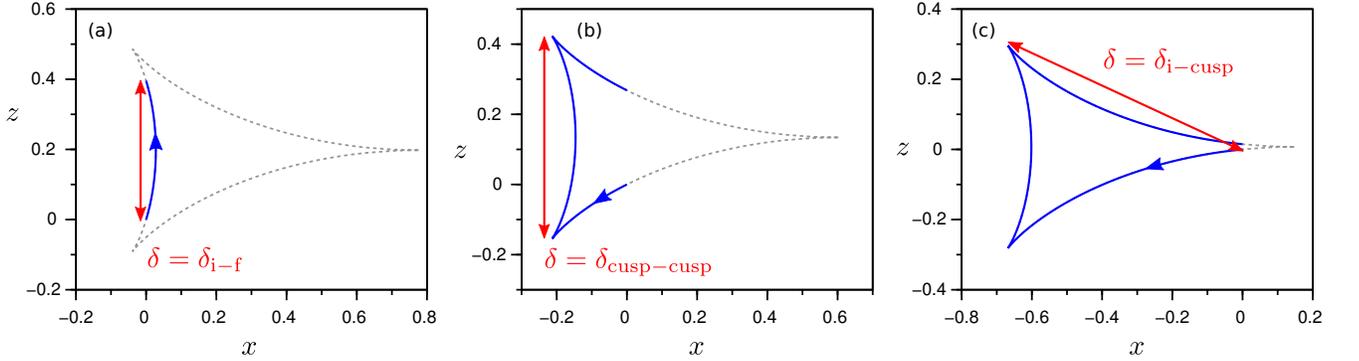}
	\caption{Paths of tracer particles, starting at the origin, advected by a dipolar swimmer moving from $x\supS=x_\mathrm{i}$ to $x\supS=-x_\mathrm{i}$ with impact parameter $\rho=1$. The path half lengths of the swimmer are (a) $x_\mathrm{i}=0.3$, (b) $x_\mathrm{i}=2$, and (c) $x_\mathrm{i}=10$. Plotted paths are the far field approximations Eqs.~\ref{eqn:farfield_xp}--\ref{eqn:farfield_zp} with $\kappa/V=3/4$. The particle range in each trajectory is indicated by a double-headed arrow. Dashed curves are the extensions of the tracer paths forward and backward in time, assuming the swimmer continues to travel along an infinite line.
	\label{fig:farfieldpaths}}
\end{figure}

For a dipole of strength $\kappa$ aligned with the swimming direction $\vec{e}$ at position $\vec{x}\supS$, the flow field is given by
\begin{equation}
\vec{u}(\vec{r})=\kappa\frac{\vec{r}}{||\vec{r}||^3}\left[3\left(\frac{\vec{e}\cdot\vec{r}}{||\vec{r}||}\right)^2-1\right],
\label{eqn:dipole}
\end{equation}
where $\vec{r}=\vec{x}-\vec{x}\supS$. According to Fax\'{e}n's laws, the velocity of a spherical particle of radius $R\supP$ in this flow field is $\vec{u}\supP=[1+(1/6)(R\supP)^2\nabla^2]\vec{u}$. To leading order in $R\supP/||\vec{r}||$, we neglect the finite size effect, treating the particle as a tracer. Simulations suggest that this is a good approximation since the particle range obtained numerically is not sensitive to $R\supP$ once $\rho>(R\supP+R\supS)$ (Fig.~\ref{fig:scaling}). The motion of the tracer is $\frac{\dd\vec{x}\supP}{\dd t} = \vec{u}(\vec{x}\supP-\vec{x}\supS)$, which can be integrated numerically over time to obtain the trajectory of the tracer. An explicit approximation can be found by considering an expansion in the particle displacement $(\vec{x}\supP-\vec{x}\supP_\mathrm{i})=\vec{x}\supP$,
\begin{equation}
\frac{\dd\vec{x}\supP}{\dd t} = \vec{u}(-\vec{x}\supS) + (\vec{x}\supP\cdot\nabla)\vec{u}(-\vec{x}\supS) +O(||\vec{x}\supP||^2/||\vec{x}\supS||^2).
\label{eqn:expand_u_dipole}
\end{equation}
Following the approach in previous studies~\cite{dunkel_swimmer-tracer_2010, pushkin_fluid_2013-1, mueller_fluid_2017}, we truncate the expansion to the zeroth order term and integrate. This yields the leading order terms in the displacement $\vec{x}\supP=(x\supP,0,z\supP)^\mathrm{T}$ of the particle,
\begin{equation}
x\supP (t)\approx \frac{-\kappa}{V} \left\{\frac{2x\supS(t)^2+\rho^2}{[x\supS(t)^2+\rho^2]^{3/2}} - \frac{2x_\mathrm{i}^2+\rho^2}{(x_\mathrm{i}^2+\rho^2)^{3/2}} \right\},
\label{eqn:farfield_xp}
\end{equation}
\begin{equation}
z\supP (t)\approx \frac{-\kappa}{V} \left\{\frac{x\supS(t)\rho}{[x\supS(t)^2+\rho^2]^{3/2}} - \frac{x_\mathrm{i}\rho}{(x_\mathrm{i}^2+\rho^2)^{3/2}}\right\}.
\label{eqn:farfield_zp}
\end{equation}
In Fig.~\ref{fig:farfieldpaths}, trajectories are plotted for a range of values of $x_\mathrm{i}$. The characteristic shapes traced out by the particles due to a dipolar swimmer have been discussed in previous studies~\cite{dunkel_swimmer-tracer_2010, pushkin_fluid_2013-1, mueller_fluid_2017}. The two cusps that appear in some of the trajectories satisfy the equation $\dd\vec{x}\supP/\dd t=\vec{0}$, which requires $x\supS=\pm\rho/\sqrt{2}$. Hence, the cusps are present only if $\rho < \sqrt{2}x_\mathrm{i}$ and have positions given by
\begin{equation}
x^\mathrm{cusp}_{\pm}\approx \frac{-\kappa}{V} \left\{2^{5/2}\cdot 3^{-3/2}\rho^{-1} - \frac{2x_\mathrm{i}^2+\rho^2}{(x_\mathrm{i}^2+\rho^2)^{3/2}} \right\},
\label{eqn:farfield_xcusp}
\end{equation}
\begin{equation}
z^\mathrm{cusp}_{\pm}\approx \frac{-\kappa}{V} \left\{\pm2\cdot3^{-3/2}\rho^{-1} - \frac{x_\mathrm{i}\rho}{(x_\mathrm{i}^2+\rho^2)^{3/2}}\right\}.
\label{eqn:farfield_zcusp}
\end{equation}

\begin{figure}[h!]
	\centering
	\includegraphics[trim = 0 0 0 0, clip, width=0.75\linewidth]{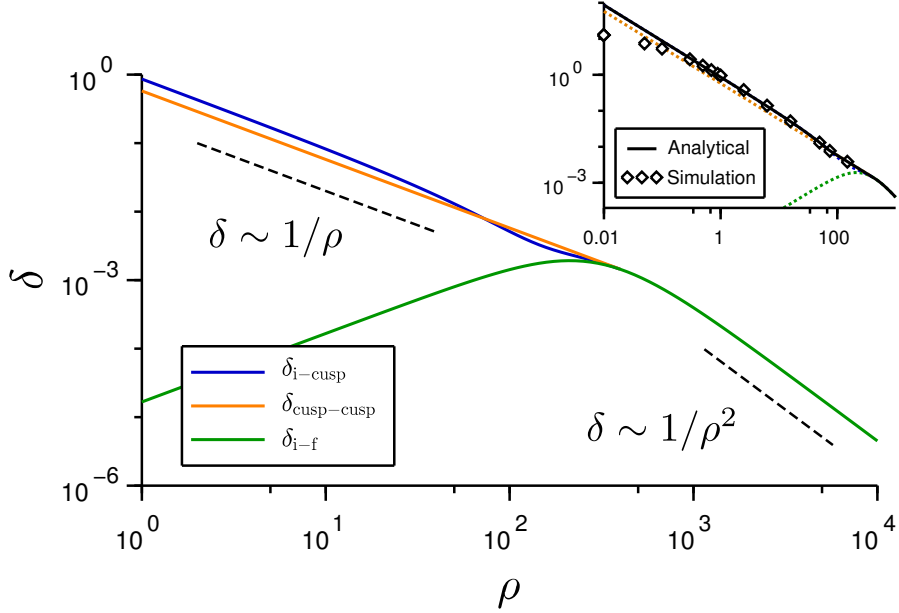}
	\caption{The dependence of particle range on impact parameter in the far field approximation. The swimmer path half length is fixed at $x_\mathrm{i}=300$. For a given impact parameter $\rho < \sqrt{2}x_\mathrm{i}$, the particle range $\delta$ is the largest of $\delta_{\mathrm{i}-\mathrm{cusp}}$, $\delta_{\mathrm{cusp}-\mathrm{cusp}}$, and $\delta_{\mathrm{i}-\mathrm{f}}$. For $\rho \geq \sqrt{2}x_\mathrm{i}$, the tracer path has no cusps and the range is $\delta=\delta_{\mathrm{i}-\mathrm{f}}$. (Inset) A comparison of $\delta$ obtained in the far field approximation with the fully simulated results for a squirmer interacting with a particle of radius $R\supP=0.01$, as in Fig.~\ref{fig:scaling}(a). Dotted curves correspond to the solid curves of the same color in the main figure. Simulations deviate from the simple, analytical results at small impact parameters, where it is not valid to neglect the near field flow around the swimmer or the finite size of the particle.
	\label{fig:farfieldrange}}
\end{figure}

The particle range is defined by
\begin{equation}
\delta = \max\limits_{0\leq t_1,t_2 \leq T_{\max}}\left\{||\vec{x}\supP(t_2)-\vec{x}\supP(t_1)||\right\},
\end{equation}
where $T_{\max}$ is the duration of the trajectory. We identify three potential candidates for maximizing the distance between points on the trajectory (see Fig.~\ref{fig:farfieldrange}):
\begin{enumerate}
\item $\delta_{\mathrm{i}-\mathrm{cusp}}$, the distance from the origin (initial position) to the second cusp,
\item $\delta_{\mathrm{cusp}-\mathrm{cusp}}$, the distance between the two cusps,
\item $\delta_{\mathrm{i}-\mathrm{f}}$, the distance from the initial to the final particle position.
\end{enumerate} 
From Eqs.~\ref{eqn:farfield_xp}--\ref{eqn:farfield_zcusp}, we obtain
\begin{equation}
\delta_{\mathrm{i}-\mathrm{cusp}}=[(x^\mathrm{cusp}_-)^2 + (z^\mathrm{cusp}_-)^2]^{1/2},
\end{equation}
\begin{equation}
\delta_{\mathrm{cusp}-\mathrm{cusp}}=|z^\mathrm{cusp}_+ - z^\mathrm{cusp}_-|=\frac{4|\kappa|}{3^{3/2}V\rho},
\end{equation}
\begin{equation}
\delta_{\mathrm{i}-\mathrm{f}}=|z\supP(T_{\max})| =\frac{|\kappa|}{V} \left\{\frac{2x_\mathrm{i}\rho}{(x_\mathrm{i}^2+\rho^2)^{3/2}}\right\}.
\end{equation}

In the limit $\rho\rightarrow\infty$ for a given, finite $x_\mathrm{i}$, there are no cusps in the particle trajectory and the range tends to
\begin{equation}
\delta=\delta_{\mathrm{i}-\mathrm{f}}\sim\rho^{-2}.
\end{equation}
In the small impact parameter regime ($\rho < \sqrt{2}x_\mathrm{i}$), the cusps must be considered when determining the particle range. The two cusps respectively minimize and maximize the $z$-coordinate of the particle. Therefore, $\delta_{\mathrm{i}-\mathrm{f}}<\delta_{\mathrm{cusp}-\mathrm{cusp}}$. In this regime, the particle range is the larger of $\delta_{\mathrm{cusp}-\mathrm{cusp}}$ and $\delta_{\mathrm{i}-\mathrm{cusp}}$. It can be shown that  $\delta_{\mathrm{cusp}-\mathrm{cusp}} < \delta_{\mathrm{i}-\mathrm{cusp}}$ whenever $\rho/x_\mathrm{i}<\rho^*$, where the threshold is at least $\rho^*\approx0.16$. Hence, for small impact parameters, the particle range is set by $\delta_{\mathrm{i}-\mathrm{cusp}}$, which has the behavior 
\begin{equation}
\delta_{\mathrm{i}-\mathrm{cusp}}=\frac{|\kappa|}{V\rho}\left\{\frac{2}{\sqrt{3}}+O(\rho/x_\mathrm{i})\right\}\sim \rho^{-1}
\end{equation}
in the limit $\rho/x_\mathrm{i}\rightarrow 0$.

To summarize, the particle range approximately exhibits two power-law scaling regimes when the swimmer path length is finite. At small impact parameters, $\delta\sim\rho^{-1}$ whereas at large impact parameters, $\delta\sim\rho^{-2}$. The crossover occurs around the threshold $\rho\approx \sqrt{2}x_\mathrm{i}$. We illustrate this behavior in Fig.~\ref{fig:farfieldrange}.

%

\end{document}